\documentclass[12pt]{article} \textheight=24 true cm
\usepackage{subfigure}
\usepackage{graphicx}
\usepackage{amsmath}
\usepackage{amssymb}
\usepackage{color}
\usepackage[colorlinks]{hyperref}
\begin{document}
\begin{center}
{\Large\bf Constraining Modified Chaplygin Gas Parameters}\\[8 mm]
 D.
Panigrahi\footnote{ Sree Chaitanya College, Habra 743268, India
\emph{and also} Relativity and Cosmology Research Centre, Jadavpur
University, Kolkata - 700032, India, e-mail:
dibyendupanigrahi@yahoo.co.in},
 B. C. Paul\footnote{Department of Physics, University of North Bengal, Dist. - Darjeeling, PIN- 734013, India,
e-mail : bcpaul@iucaa.ernet.in}
  and S. Chatterjee\footnote{IGNOU Convergence
Centre, New Alipore College, Kolkata - 700053, India \emph{and
also} Relativity and Cosmology Research Centre, Jadavpur
University, Kolkata - 700032,India
e-mail : chat\_ sujit1@yahoo.com} \\[6mm]

\end{center}

\begin{abstract}
We investigate the evolution of a FRW model fuelled by a modified
Chaplygin gas with an equation of state $p = A\rho
-\frac{B}{\rho^\alpha}$.  An attempt is made here to constrain the
free parameters of MCG model through the wellknown contour plot
technique using observational data. The permissible range of
values of the pair of free parameters are determined to  study the
viability of cosmological models \emph{vis-a-vis} observational
results. Aside from allowing the desirable feature of \emph{flip}
of the sign of deceleration parameter  we also find that the
transition from the decelerating to the accelerating phase occurs
at relatively low value of redshift in accordance  with the
observational prediction that the acceleration is a recent
phenomenon. Stability of the model against density perturbation is
studied in some detail and it is found that the effective acoustic
speed may become imaginary depending upon the initial conditions
signalling that perturbations associated with instability sets in
resulting in structure formation. As one considers more negative
values of $A$ the \emph{flip}  in sign is delayed causing  the
density parameter to change fast. Again it is found from the
contour plot that compatibility with observational results is
admitted with a value of $A$ which is very near to zero or a small
negative number.

\end{abstract}

\vspace{0.5 cm} ~~~~~~ PACS :   98.80.-k,98.80.Es
\bigskip
\section{ Introduction}

 Following the high redshift supernovae data in
the last decade~\cite{riess} we know that when interpreted within
the framework of the standard FRW type of universe (homogeneous
and isotropic) we are left with the only alternative that the
universe is now passing through an accelerated phase of  expansion
with baryonic matter contributing only four percent of the total
energy budget. Observational results coming  from CMBR studies
~\cite{wmap} also point to this conclusion engaging a large
community of cosmologists ~\cite{saif} and references therein) to
embark on a quest to explain the cause of the acceleration. The
teething problem now confronting researchers is the identification
of the mechanism that triggered the late inflation. Researchers
are mainly divided into two groups � either suggesting a
modification of the original general theory of relativity or
invoking any mysterious fluid in the form of an evolving
cosmological constant or a quintessential type of scalar field.
But as debated at length in the literature both the alternatives
face serious problems mainly at the theoretical and conceptual
level.
 Typically, the late acceleration  has been attributed
to a mysterious dark energy (DE) which contributes about three
fourth of the cosmic substratum. Another unknown matter component
of the universe is the dark matter (DM), the missing mass
necessary to hold together the galaxy clusters and also needed to
explain the current large scale structure of the universe.
\vspace{0.5 cm}

In the literature  a good number of dark energy models are
proposed but little is precisely known about it. Nowadays, the
dark energy problem remains one of the major unsolved problems of
theoretical physics~\cite{wen}. On the way of searching for
possible solutions of this problem various models are explored
during last few decades referring to e.g. new exotic forms of
matter (\emph{e.g.} quintessenceï) ~\cite{quint}, phantom
~\cite{phan}, holographic models ~\cite{hol}, string theory
landscape~\cite{string} Born-Infeld quantum condensate
~\cite{quantum}, modified gravity approaches~\cite{star},
inhomogeneous spacetime~\cite{krasin} etc~ (readers interested in
more detail for a comprehensive overview of existing theoretical
models may refer to~\cite{add})
 \vspace{0.5 cm}

Even though the cosmological constant allows for the cosmic
acceleration at late times, the observational bounds on $\Lambda$
are incompatible with theoretical predictions of a gravitational
vacuum state. Grosso modo, the $\Lambda$ model does well in
fitting most observational data but the density parameter
corresponding to  $\Lambda$ and matter are of the same order of
magnitude,  surprisingly close to each other,  even though
$\Lambda$ is a   constant during the entire evolutionary history
of the universe. These two thorny shortcomings, namely the fine
tuning and the coincidence problems, disturb the otherwise
appealing picture of a cosmological constant and dramatically
plague the so called $\Lambda$CDM model. To circumvent this
difficulty and host others people invoke a variable $\Lambda$
model but that too is beset with serious field theoretic problems
~\cite{lamda}. Another possibility to obtain accelerated expansion
is provided by theories with large extra dimensions known as
braneworlds ~\cite{sc} but one has to hypothesize the existence of
extra spatial dimensions in these models. \vspace{0.5 cm}

      Among the different theories put
forward in the literature in recent times, the single component
fluid known as Chaplygin gas(CG) with an equation of state (EoS)
$p = -\frac{B}{\rho}$ ~\cite{cg}, where $\rho$ and $p$ are the
energy density and pressure respectively and $B$ is a constant has
attracted large interest in cosmology. Although the model is very
successful in explaining the SNe Ia data it shows that CG does not
pass the tests connected with structure formation and observed
strong oscillations of matter power spectrum ~\cite{sand}. One can
circumvent this situation  in the generalised chaplygin gas(GCG)
proposed with $p=-\frac{B}{\rho^{\alpha}}$ with $\alpha$,
constrained in the range $0<\alpha < 1$. Since the inferences from
the $\Lambda CDM$ and GCG models are almost similar, atleast from
the observational fallouts this model is further modified either
considering $B$ as a function of the redshift($z$) as $B=B(z)$
~\cite {zhang} or assuming an EoS as,
\begin{equation}
p = A\rho -\frac{B}{\rho^{\alpha}}
\end{equation}
(MCG)~\cite{ben}, where $A$ is a new constant parameter of the
theory. Thus MCG may be looked upon as a mixture of a barotropic
perfect fluid and GCG.  Although it suffers from serious
shortcomings e.g., it violates the time honoured principle of
energy conditions, still its theoretical conclusions are found to
be in broad agreement with the observational results coming out of
gravitational lensing or recent CMBR and SNe data in varied cosmic
probes. This is generally achieved through a careful maneuvering
of the value of the newly introduced  arbitrary constant. The
viability of the scenario has been tested by a number of
cosmological probes, including SNe Ia data~\cite{fabris}, lensing
statistics~\cite{deb}, age-reshift tests~\cite{alcaniz}, CMB
measurements~\cite{bento}, measurements of X Ray luminosity of
galaxy clusters~\cite{cunha}, statefinder parameters~\cite{sahni}.
Another interesting feature of the MCG is that it can show
radiation era in the early universe. At late times MCG behaves as
a cosmological constant and can be fitted to $\Lambda CDM$.
\vspace{0.5 cm}

 Furthermore,  dark matter (in short,  DM) and dark energy (DE)
are considered as different manifestations of the same component
and describes the dark sector as some kind of fluid whose physical
properties depend on the scale: it behaves as DM at high densities
and transforms into DE at lower ones. Most of these United Dark
Matter models invoke the generalized Chaplygin gas (GCG), a
perfect fluid characterized by a negative pressure which is
inversely proportional to the energy density.
\vspace{0.5 cm}

Using modified chaplygin gas (in short, MCG) in FRW universe a
number of cosmological models  have been discussed in the
literature. In most of the models  the values of the so called
free parameters are picked up by hand  in order to suit the best
fit values with the observational results without much physical
considerations. It is important to study cosmological models from
observational contexts  which naturally impose a constraint on the
values of the free parameters. In view of the above
considerations, in this work  we determine the constraints on the
free parameters of the model by drawing the usual contour plot
diagrams. However two works in this field are worth mentioning. In
an earlier one Jianbo Lu \emph{et al}~\cite{lu}have apparently
come to the conclusion via usual contour plot diagram that the
recently observed data give credence to both the MCG as well as
other fashionable models so that it is very difficult to choose
one over the other. This is in line with earlier results in this
field~\cite{steinhardt}. Secondly in a very recent communication
Wang \emph{et al}~\cite{wang} have also studied in depth the
constraints on $\alpha$ in GCG formalism in two distinct cases -
one pertaining to a barotropic equation of state and the other to
a dark matter with some specific properties. In what follows we
have discussed and compared, in brief, our own findings in this
field \emph{vis-a-vis}     Wang et al's work at the end of our
paper in sec. 5.  Our work is primarily focussed on two domains.
First to study the dynamics of FRW cosmological scenario with MCG
as matter field  as also to obtain constraints  on the so called
three free parameters with the help of contour diagram to achieve
the best fit with current observational findings. The paper is
organised as follows: in sec. 2 we build up the evolution
equations, while the nature of evolution and analysis of
instability of our model are dealt with in some detail in sec. 3.
In sec. 4 we discuss the observational data, namely Stern data,
measurement of baryon acoustic oscillations (BAO) peak parameters
and CMB shift data to draw contour diagram for the permissible
range of values of the pair of parameters ($\alpha ,
A$),($\omega_{ode} , A$) and ($\Omega_{odm}, A$). Finally in sec.
5  we give a brief discussion.

\section{ Field Equations}

We consider a flat spherically symmetric homogeneous spacetime
with line element
\begin{equation}\label{a}
  ds^{2}= dt^{2}- a^2(t)~(dr^{2}+r^{2}d\Omega^{2})
\end{equation}
where the scale factor, $a(t)$  depends on time only. Taking a
comoving coordinate system such that $ u^{0}=1, u^{i}= 0 ~(i = 1,
2,3)$ and $g^{\mu \nu}u_{\mu}u_{\nu}= 1$ where $u_{\mu}$
represents the 4- velocity, the energy momentum tensor for a
perfect fluid distribution in the above defined coordinates is
given by
\begin{equation}
T^{\mu}_{\nu} = (\rho + p)\delta_{0}^{\mu}\delta_{\nu}^{0} -
p\delta_{\nu}^{\mu}
\end{equation}
where $\rho(t)$ is the matter density  and $p(t)$ the isotropic
 pressure.
The Einstein field equations  for the above metric are given by
\begin{eqnarray}
3 H^{2} &=& \rho \\
2\dot{H} + 3 H^{2} &=& -p
\end{eqnarray}
where $H (= \frac{\dot{a}}{a})$ represents the Hubble parameter
and we use units where 8 $\pi \;G$ = 1. The conservation equations
for matter fields are given by
\begin{equation}
\nabla_{\nu}T^{\mu \nu}= 0
\end{equation}
which in turn yields
\begin{equation}
\dot{\rho}_{total} + 3 H (\rho_{total} + p_{total}) = 0
\end{equation}
We now consider a mixture of fluids, where baryons represents one of the component
 and conservation of baryons leads to the continuity equation that  can be
 separated into two equations, given by
\begin{eqnarray}
\dot{\rho}_{b} + 3 H (\rho_{b} + p_{b}) = 0 \\
\dot{\rho} + 3 H
(\rho + p) = 0
\end{eqnarray}
where $\rho_{total} = \rho + \rho_{b}$ and the subscript $b$
denotes the baryons. The component of other part is a mixture of dark matter and dark energy.
 In our model we consider Modified Chaplygin gas (MCG) as one of the candidate for dark energy
 and consequently the mathematical expressions are  given by   $p = p_{de}$ and
$\rho = \rho_{de} +\rho_{dm}$, which obeys an EoS given by
equation (1). In the above, the subscripts $dm$ and $de$ denote
dark matter and dark energy respectively. Let
  $ p_{de}=\omega_{de}\rho_{de} $, where  the EoS parameter  for dark
energy is $\omega_{de}$. Using eq. (1) in conservation equation
given by eq. (9), one obtains  an energy density  given by
\begin{equation}
\rho = \left[ \frac{B}{1+A} + c(1+z)^{3(1+\alpha)(1+A)}\right
]^{\frac{1}{1+\alpha }}
\end{equation}
where $c$ is an integration constant.  From eq.  (10)  it is evident that $A
\neq -1$ is required for a finite  $\rho$ .

 \vspace{0.1 cm}
\section{Cosmological Dynamics}

\begin{figure}[ht]

\begin{center}
  \includegraphics[width=8cm]{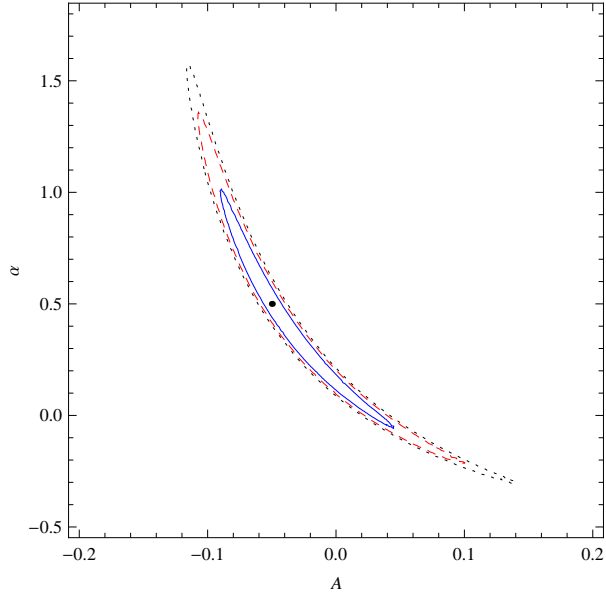}
  \caption{
  \small\emph{The contour for $\alpha$ and $A$  for $\Omega_{odm} = 0.22$, $\omega_{ode} = - 1.05$.  }\label{1}
    }
\end{center}
\end{figure}

The energy densities of the dark energy and the dark matter are
now determined using  the set of dynamical equations given in the
sec. 2 such that
\begin{eqnarray}
\rho_{de} = \frac{A
\left[\frac{B}{1+A}+c(1+z)^{3(1+\alpha)(1+A)}\right]-B}{\omega_{de}\left[
\frac{B}{1+A} + c(1+z)^{3(1+\alpha)(1+A)}\right
]^{\frac{\alpha}{1+\alpha }}},
\end{eqnarray}
\begin{eqnarray}
\rho_{dm} = \rho - \rho_{de} = \frac{\left( \omega_{de} - A
\right) \left[\frac{B}{1+A} + c(1+z)^{3(1+\alpha)(1+A)} \right]
+B}{\left[ \frac{B}{1+A} + c(1+z)^{3(1+\alpha)(1+A)}\right
]^{\frac{\alpha}{1+\alpha }}}. \hspace{1.8 cm}
\end{eqnarray}
Now the constants $B$ and $c$ can be expressed as,
\begin{equation}
B = \left \{A(1-\Omega_{ob}) - \omega_{ode}\Omega_{ode} \right
\}(1- \Omega_{ob})^{\alpha}\rho_{oc}^{(1+\alpha)}
\end{equation}

\begin{equation}
c = (1-\Omega_{ob} + \omega_{ode}\Omega_{ode} )\frac{(1-
\Omega_{ob})^{\alpha}}{1+A}\rho_{oc}^{(1+\alpha)}
\end{equation}
where $\rho_{oc}$ represents  critical density, and label $o$
denotes quantities that estimate present values.
 \vspace{0.5 cm}

 In the case of  GCG, $B$ is one
free parameter but in the Modified Chaplygin gas (MCG) the
parameter $B$ is related to $A$ and the present value of the
density parameter, $\alpha$ and the EoS parameter via equation
(13) so that when we constrain $A$ from the observational inputs
the other constant $B$ is automatically fixed up. Further from the
equation (10) we find that $B$ is important when $z$ is very small
and the universe is dark energy dominated. So knowing the dark
energy contribution $B$ can be fixed, which, however, is not
considered in the present work.  \vspace{0.5 cm}

\begin{figure}[ht]

\begin{center}
  \includegraphics[width=8cm]{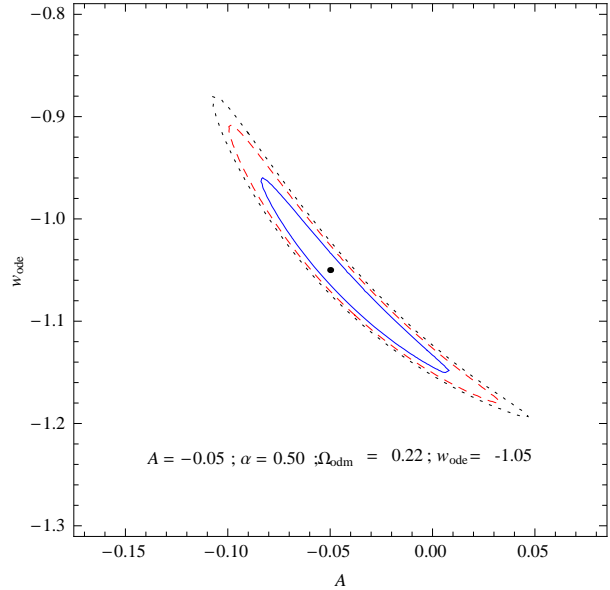}
  \caption{
  \small\emph{The contour for  $\omega_{0de}$ and  $A$  is shown.}\label{1}
    }
\end{center}
\end{figure}

\begin{figure}[ht]

\begin{center}
  \includegraphics[width=8cm]{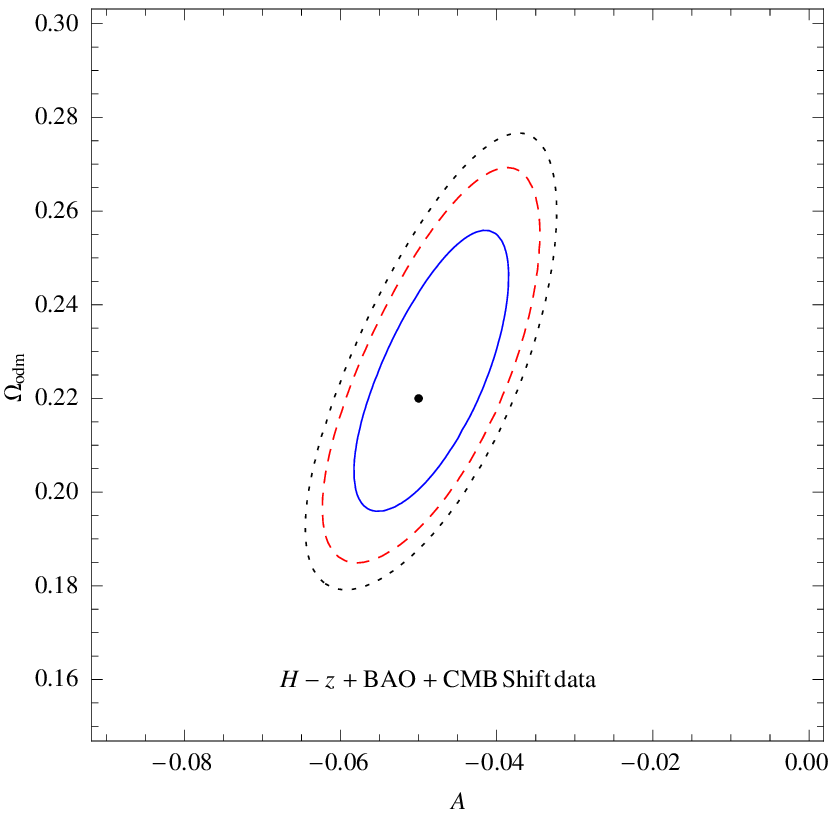}
  \caption{
  \small\emph{The contour for $\Omega_{0dm}$ and $A$  is shown.  }\label{1}
    }
\end{center}
\end{figure}

 The energy density of the universe can be expressed in terms of redshift
 parameter\emph{,} $z$ which is given by
\begin{multline}
\rho(z)= \frac{(1 -
\Omega_{ob})^{\frac{\alpha}{1+\alpha}}}{(1+A)^{\frac{1}{1+\alpha}}}
\biggl[A(1-
\Omega_{ob}) - \omega_{ode}\Omega_{ode} \\
+ (1 - \Omega_{ob}  +
\omega_{ode}\Omega_{ode})(1+z)^{3(1+\alpha)(1+A)}
\biggr]^{\frac{1}{1+\alpha}}
\end{multline}
where we use $a (t) = \frac{1}{1+z}$.

\begin{figure}[ht]
\begin{center}
  \includegraphics[width=8cm]{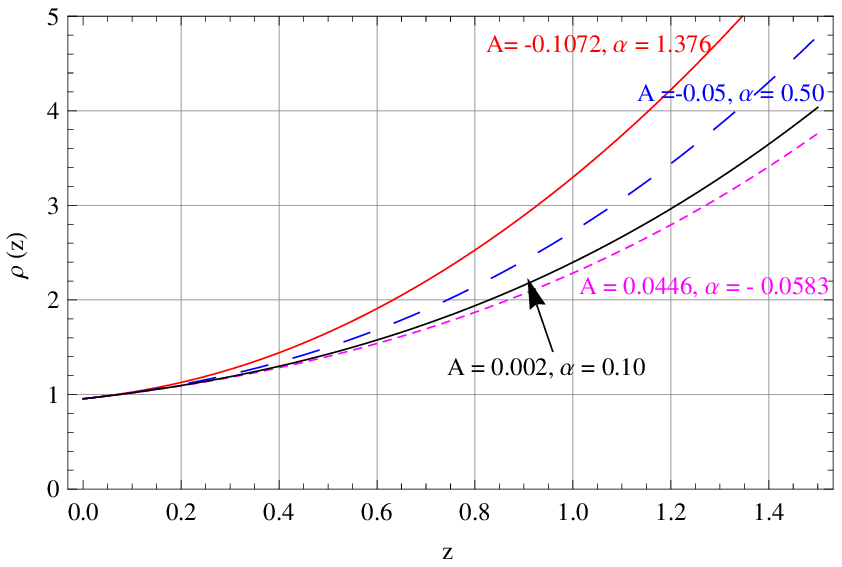}
  \caption{
  \small\emph{The densities  as
  the function of redshift. Taking $\omega_{ode} = -1.05$,
   $\Omega_{ode} = 0.734$, $\Omega_{odm} = 0.222 $ and $\Omega_{ob}= 0.044$.  }\label{1}
    }
\end{center}
\end{figure}

 \vspace{0.5 cm}
We consider four values of $A$ \& $\alpha$ as permitted by the
contours drawn  in figs. (1) \& (2),  the density variation with
$z$  are plotted and found that the universe is  denser for $A =
-0.1072 $ and $\alpha = 1.376$. Other permitted values of $A$ and
$\alpha$ are $(- 0.05, 0.50); (0.002, 0.100);(0.0446, - 0.0583)$
respectively. However, from fig. (3) it is clear that  the range
of $A$ is further restricted to ($-0.0649 $ to $- 0.0319$)  at
$95.4 \% $ confidence level. This implies that the dark matter
parameter dictates
 that $A$ should be slightly less than zero. The prediction in our model
 agrees with that
 obtained by   Fabris \emph{et al}~\cite{fab3}. Using the WMAP predicted
  values \cite{wmap} namely, the current
density parameters  $\Omega_{ode} = 0.734 \pm 0.029$,
$\Omega_{odm} = 0.222 \pm 0.026 $ and $\Omega_{ob}= 0.0449 \pm
0.0028$, we explore the variation of the density of the universe
$\rho(z)$ with redshift  $z$. The variation of $\rho(z)$ with $z$
is plotted in fig. (4) for different $A$ and $\alpha$ values. As
expected the  density of the universe increases with an increase
in $z$ values, as more
 the value of $z$ more the observations probe the early universe. It is also
  noted that as one picks up more and more negative  values of
  $A$ with positive $\alpha$, it leads to greater energy density in the early
  universe for a fixed $z$ value. The permitted range of values of $A$ and $\alpha$ will
  be discussed in sec. 4, the  present energy density of the universe is
independent of  $A$ \& $\alpha$ at $z = 0$ and attains a value
$(\rho)_{z=0} = (1 - \Omega_{ob})$ which also follows from the
eq.(15). The study reveals the dynamical property of the universe
which shows how fast the universe is accelerating for different
values of $A$ \& $\alpha$.
\vspace{0.5 cm}

Using eqs.  (1), (10), (13) and (14), we obtain the  equation of
state parameter for the MCG model given by

\begin{equation}
\omega(z)=  \frac{p}{\rho} = - \frac{A(1- \Omega_{ob}) -
\omega_{ode}\Omega_{ode}-A(1- \Omega_{ob} +
\omega_{ode}\Omega_{ode})(1+z)^{3(1+\alpha)(1+A)}}{A(1-
\Omega_{ob}) - \omega_{ode}\Omega_{ode}+(1- \Omega_{ob} +
\omega_{ode}\Omega_{ode})(1+z)^{3(1+\alpha)(1+A)}}
\end{equation}

For consistency check we note that with $B= 0$ eq. (13) gives
$A= \frac{\omega_{0de}\Omega_{ode}}{1- \Omega_{ob}}$ leading to a
barotropic equation of state parameter $\omega(z) = A$,  which is
negative as  $\omega_{ode}$ is  negative definite. In our case an   accelerating universe is permitted
with $A < -\frac{1}{3}$. Considering $\omega_{ode} = -1.05$,
   $\Omega_{ode} = 0.734$, $\Omega_{odm} = 0.222 $ and $\Omega_{ob}=
   0.044$, we get $A \approx - 0.80$ for  a viable cosmology. This model is in agreement with
     $\Lambda$CDM model.  For $A = 0$, this MCG model reduces to
GCG model. In this case
\begin{equation}
\omega(z) =
\frac{\omega_{ode}\Omega_{ode}}{-\omega_{ode}\Omega_{ode} +
\left(1- \Omega_{ob} + \omega_{ode}\Omega_{ode}
\right)(1+z)^{3(1+\alpha)}}
\end{equation}
It is noted from the contour drawn in fig. (2) that $\omega_{ode}$
lies between $- 1.194$ to $- 0.8771$ at $95.4 \%$ confidence
limit.  It is also noted that the function $\omega(z)$ is always
negative for entire range of values of $\omega_{ode}$ with (i) $A=
- 0.05$ \& $\alpha =0.50$, (ii) $A= - 0.1072$ \& $\alpha =1.376$
and (iii) $A= 0.002$ \& $\alpha = 0.100$. However, we note that
$\omega (z)$ is negative at the present epoch
 which was positive in the early epoch for
 $A=0.0446$ and $\alpha =- 0.0583$ . Thus a transition
 to an accelerating universe is permitted.
 \vspace{0.5 cm}

In fig. (1),  we plot variation of $A$ with $ \alpha$. It is
evident that it admits
 $A = 0$, $ \alpha = 0.10$. At $z = 0$, $\omega(z)$ becomes
negative for negative values of $\omega_{ode}$, admitting an
accelerating universe. The evolution of $\omega(z)$ with $z$ is
shown in the fig. (5). It is evident   that $\omega(z)$ is mostly
flat in the high redshift region ($z>2.0$) which is very steep at
low redshift ($z$) region. We note that $\omega(z)$ is always
negative as one extrapolates from past to future for the values of
$A = - 0.1072$ \& $\alpha = 1.376$ , $A = - 0.05$ \& $\alpha =
0.50$ and $A= 0.002$ \& $\alpha = 0.100$, but for $A = 0.0446$ \&
$\alpha = - 0.0583$,  it is positive in the past (shown with   $z
\approx 3.80$). In this case we get $\omega(z) = 0$ at $z \approx
3.80$, corresponding to a dust dominated universe in the recent
past. Again it is found that at $z = 0$, $\omega(z)$ attains
$-0.80$ which is independent of $A $ \& $ \alpha$. It also follows
from the eq. (16) that $\omega(z)|_{z=0} =
\frac{\omega_{ode}\Omega_{ode}}{1 - \Omega_{ob}} $.

\begin{figure}[ht]

\begin{center}
  \includegraphics[width=8cm]{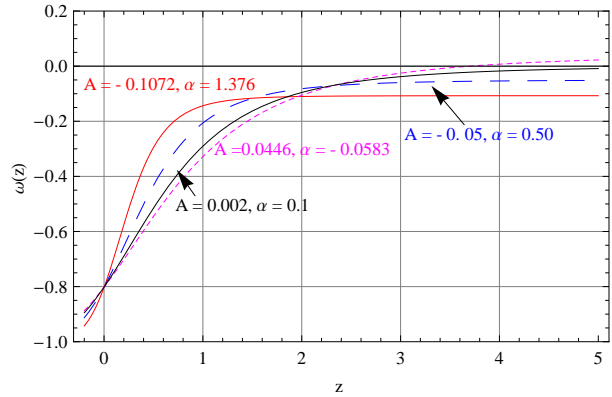}
  \caption{
  \small\emph{$\omega$(z) as
  the function of redshift. Taking $\omega_{ode} = -1.05$,
   $\Omega_{ode} = 0.734$, $\Omega_{odm} = 0.222 $ and $\Omega_{ob}= 0.044$.   }\label{1}
    }
\end{center}
\end{figure}

In the context of the theory of linear perturbation the squared
speed of sound, $v_{s}^{2}$, plays a very important role in
determining the stability or instability of a given perturbed
state. The positive sign indicates the periodic propagating mode
for a density perturbation and the system points to stability for
a given mode. The negative sign (imaginary value of speed) shows
an exponentially growing mode for a density perturbation
indicating the instability of a given mode. Generally the
evolution of sound speed in the linear regime of perturbation is
independent of the dynamics of the background cosmology as
follows:
\begin{equation}
v_{s}^{2}= \frac{dp}{d\rho}= \frac{1}{3H}\frac{d}{dH}\left[H^{2}
\left(q - \frac{1}{2}\right)\right]
\end{equation}
Thus the sign of $q$ determines the sign of $v_{s}^{2}$ via the
transition epoch from the CDM dominated phase to DE dominated
phase. From the above description one can easily find that the
growth of perturbation is dependent on the choice of DE model of
background dynamics. For the Chaplygin gas (CG) model we have
\begin{equation}
v_{s}^{2}= -\frac{B}{\rho^{2}}= -\omega(z)
\end{equation}
If we further take  the CG as a quintessence DE model with $-1
<\omega(z) < 0$, the squared speed is always positive. So the
stability of the CG model against density perturbation at any
 cosmic scale factor is guaranteed.\\
In the case of GCG model the situation is almost similar. We see
\begin{equation}v_{s}^{2} = -
\alpha\omega(z)
\end{equation}
Once again in the quintessence range the GCG model is instable
against density perturbation if $\alpha <0$ and stable if $\alpha
> 0$. From the observational data we know that $\alpha$ is always
positive and this ensures the stability of the model. For the MCG
model under consideration the situation is, however more involved.
We get for quintessence zone
\begin{equation}
v_{s}^{2}= A(1 + \alpha)- \alpha \omega(z)
\end{equation} Thus for $A \geq 0$ the system
is always stable. But for negative $A$ the stability depends on
the relative magnitudes of $\omega(z)$, $A$ and $\alpha$.

Now  eq.,(16) and(21) further give for our case under
consideration

\begin{eqnarray}
v_{s}^{2} =  A(1+\alpha)+ \alpha \left[\frac{A(1- \Omega_{ob}) -
\omega_{ode}\Omega_{ode}-A(1- \Omega_{ob} +
\omega_{ode}\Omega_{ode})(1+z)^{3(1+\alpha)(1+A)}}{A(1-
\Omega_{ob}) - \omega_{ode}\Omega_{ode}+(1- \Omega_{ob} +
\omega_{ode}\Omega_{ode})(1+z)^{3(1+\alpha)(1+A)}} \right]
\end{eqnarray}

\begin{figure}[ht]
\begin{center}
  \includegraphics[width=8cm]{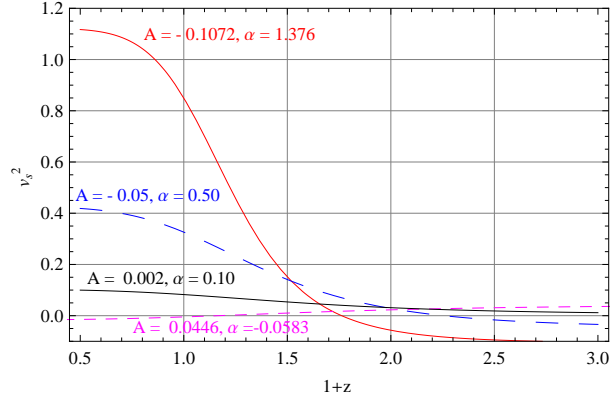}
  \caption{
  \small\emph{The squared speed of sound vs redshift is shown in this figure. Taking $\omega_{ode} = -1.05$,
   $\Omega_{ode} = 0.734$, $\Omega_{odm} = 0.222 $ and $\Omega_{ob}= 0.044$.  }\label{1}
    }
\end{center}
\end{figure}

In fig.(6) we plot the variation of the squared speed of sound
$v^{2}_{s}$ with redshift parameter $z$. It is observed that the
speed of sound $v_{s}$ is positive definite for $A = - 0.1072$ \&
$ \alpha = 1.376$  and $A = - 0.05$ \& $ \alpha = 0.50$
(\emph{i.e.}, negative values of $A$) for low value of $z$, which,
however, becomes $v_{s}^{2} < 0$ for high redshift region. This
leads to instability of the MCG fluids as noted  by Mei Deng
~\cite{mei}. It is argued that for $\alpha
> -1$, $(v_{s}^{2})_{z\rightarrow \infty} \approx 0$. However,
we note that instability  sets in our case with $\alpha \geq
-0.0583$ leading to  $v_{s}^{2} = 0$ for finite $z$ accommodating
the results obtained by Mei Deng.
\vspace{0.5 cm}

We note that the squared speed of sound does not exceed that of
light for $A = - 0.1072$ \& $ \alpha = 1.376$ at $z = 0$. However,
in future, \emph{i.e.}, ($1+z <1$) it is possible, $v_{s}$ exceeds
the speed of light for the above value of $A$ and $\alpha $. This
results in a perturbation of the spacetime and a perturbative
analysis of the whole system shows that it favours structure
formation~\cite{fab}.  For $A = 0.0446$ \& $ \alpha = - 0.0583$,
the MCG fluids are unstable for lower $z$ value.  However for high
redshift, i.e., in the early universe $v_{s}^{2} > 0$ is
permitted. It is very interesting to note that for $A= 0.002$ \&
$\alpha = 0.100$, $v_{s}^{2}$ is always positive. At high redshift
region it was lower than present epoch. Again in future
$v_{s}^{2}$ will be higher but does not exceed the velocity of
light. At $A = 0$, $v_{s}^{2} = -\alpha \omega(z)$ and $\omega(z)
<0$ in this case, which gives positive value of $v_{s}$(since
$\alpha$ is slightly greater than zero at $A = 0$). From eq. (22)
it follows that the present value of $v_{s}^{2}$ is
$(v_{s}^{2})_{z=0} = A (1+\alpha) - \frac{\alpha
\omega_{ode}\Omega_{ode}}{1 - \Omega_{ob}}$ obtained by putting $z
= 0$. The squared  speed of sound $v_{s}^{2}$ depends both on $A$
and $\alpha$ at low redshift $z = 0$. This is unlike the cases of
$\rho$ and $\omega(z)$.
\vspace{0.5 cm}

 It is evident  from fig-6 that at
high redshift region ( $ 1 + z < 1.7$ ) $v_{s}^{2} < 0$ for  $A =
-0.05, \alpha = 0.50$ and  $A = - 0.1072, \alpha = 1.376$. This
leads to instability of the MCG fluid at high $z$, i.e., at early
universe, which helps in structure formation. But the model is
stable at the present epoch (i.e., $z = 0$) and also $ 0 <
v_{s}^{2} < c_{s}^{2}$.
 \vspace{0.5 cm}

 Now, we   consider a spatially flat FRW universe with  background fluid
  described by an exotic fluid component
namely, MCG and baryon component. From the dynamical field
equations we obtain dimensionless Hubble parameter
\begin{multline}
E^{2}(z) \equiv \frac{H^{2}}{H_{o}^{2}}
 = \frac{(1 -
\Omega_{ob})^{\frac{\alpha}{1+\alpha}}}{(1+A)^{^{\frac{\alpha}{1+\alpha}}}}
 \biggl[A \left(\Omega_{ode}+ \Omega_{odm} \right)-
\omega_{ode}\Omega_{ode} \\
+ \Omega_{odm}(1+z)^{3(1+\alpha)(1+A)}+
\Omega_{ode}(1+\omega_{ode})(1+z)^{3(1+\alpha)(1 + A)}
\biggr]^{\frac{1}{1+ \alpha}} + \Omega_{ob}(1+z)^{3}
\end{multline}

\begin{figure}[t]
\begin{center}
  \includegraphics[width=8cm]{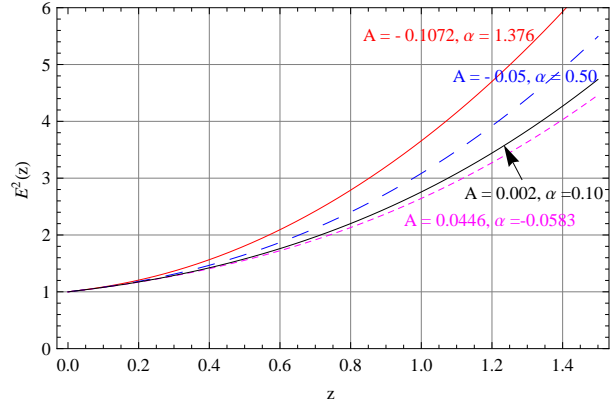}
  \caption{
  \small\emph{The variation of Hubble parameter vs $z$  is shown in this
  figure.  Taking $\omega_{de} = -1.05$,
   $\Omega_{ode} = 0.734$, $\Omega_{odm} = 0.222 $ and $\Omega_{ob}= 0.044$.   }\label{1}
    }
\end{center}
\end{figure}

The variation of normalized Hubble parameter with redshift is
shown in the fig. (7). It is observed that normalised Hubble
parameter changes more sharply with $z$ for the set ($A= -0.1072$,
$\alpha =1.376$) compared to other sets, where $A$ is less
negative in line with the fact that more negative $A$ values imply
a more accelerating model. It attains   $E^{2}(z) =1$ at the
present epoch $z = 0$.

Using the above equations, the densities of different components
of the energy namely, $\Omega_{dm}$, $\Omega_{de}$ and
$\Omega_{b}$ are obtained which are given below :

\begin{multline}
\Omega_{de}
 =
 \frac{(1-\Omega_{ob})^{\frac{\alpha}{1+\alpha}}(1+A)^{\frac{\alpha}{1+\alpha}}}{\omega_{de}E^2(z)}
 \biggl[\frac{A}{1+A}
  \biggl\{A(1-\Omega_{ob})-\omega_{ode}\Omega_{ode}\\
 +(1-\Omega_{ob}+
 \omega_{ode}\Omega_{ode})(1+z)^{3(1+\alpha)(1+A)} \biggr\}^{\frac{1}{1+\alpha}}  \hspace{4.5cm}
 \\
 - \frac{A(1-\Omega_{ob}) -
 \omega_{ode}\Omega_{ode}}{\{A(1-\Omega_{ob}) - \omega_{ode}\Omega_{ode} + (1-\Omega_{ob}+
 \omega_{ode}\Omega_{ode})(1+z)^{3(1+\alpha)(1+A)}\}^{\frac{\alpha}{1+\alpha}}}\biggr]
\end{multline}

\begin{multline}
\Omega_{dm}
 =
 \frac{(1-\Omega_{ob})^{\frac{\alpha}{1+\alpha}}(1+A)^{\frac{\alpha}{1+\alpha}}}{\omega_{de}E^2(z)}
 \biggl[\frac{1}{1+A}(\omega_{de}-A)
 \biggl\{ A(1-\Omega_{ob})-\omega_{ode}\Omega_{ode}  \\
 +(1-\Omega_{ob}+
 \omega_{ode}\Omega_{0de})(1+z)^{3(1+\alpha)(1+A)}\biggr\}^{\frac{1}{1+\alpha}} \hspace{4.5cm}
 \\
 + \frac{A(1-\Omega_{ob}) -
 \omega_{ode}\Omega_{ode}}{\{A(1-\Omega_{ob}) - \omega_{ode}\Omega_{ode} + (1-\Omega_{ob}+
 \omega_{ode}\Omega_{ode})(1+z)^{3(1+\alpha)(1+A)}\}^{\frac{\alpha}{1+\alpha}}}\biggr]
\end{multline}

\begin{figure}[ht]
\begin{center}
  \includegraphics[width=8cm]{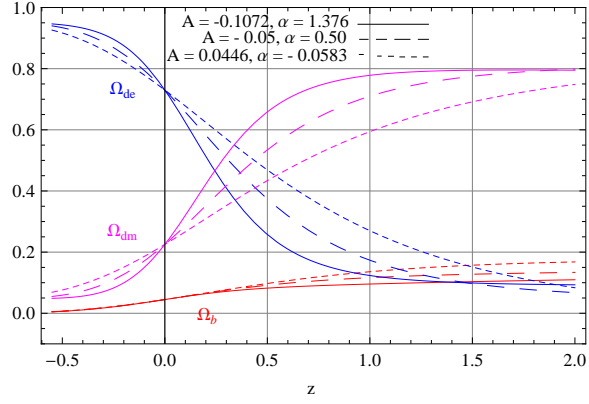}
  \caption{
  \small\emph{The densities of $\Omega_{de}$,$\Omega_{dm}$ and $\Omega_{b}$ as
  the function of redshift. Taking $\omega_{ode} = -1.05$,
   $\Omega_{ode} = 0.734$, $\Omega_{odm} = 0.222 $ and $\Omega_{ob}= 0.044$.    }\label{1}
    }
\end{center}
\end{figure}
and

\begin{equation}
\Omega_{b} = \frac{\Omega_{ob}(1+z)^{3}}{E^{2}(z)}.
   \hspace{4.5cm}
\end{equation}

The variation of density parameters $\Omega_{de} $, $\Omega_{dm} $
and $\Omega_{b}$ with the redshift $z$ is shown in the fig (8). At
$z = 0$, the values of  the density parameters attain the  same
value for different sets of the values of $A$ \& $\alpha$. For
example for the set ($A = - 0.1072$, $ \alpha = 1.376$) density
parameters change with $z$ at a  very fast rate than the other
sets of the values of $A$ \& $\alpha$ taken up here. It is shown
from the fig. (8)  that the change of the density parameters is
faster for more negative values of $A$ admitting an accelerating
universe. It is evident from  fig. (8) that the dominance of the
dark energy leads to the acceleration. The gradual increasing
density of dark energy is the cause of the expansion of the
universe which transits from deceleration  to acceleration phase.
To understand this transition we study the evolution of the
deceleration parameter $q$.

Now the deceleration parameter is given by,

\begin{multline}
 q \equiv -1 - \frac{\dot{H}}{H^{2}}\\
= -1 + \frac{3(1+z)^{3}}{2E^{2}(z)}
\biggl[(1+A)^{\frac{\alpha}{1+\alpha}}(1-\Omega_{ob})^{\frac{\alpha}{1+\alpha}}(1+z)^{3A}(1-\Omega_{ob}+
 \omega_{ode}\Omega_{0de})\\
\biggl \{ \{A(1-\Omega_{ob}) -
 \omega_{ode}\Omega_{ode}\}(1+z)^{-3(1+\alpha)(1+A)} + 1-\Omega_{ob}+
 \omega_{ode}\Omega_{ode})\biggr \}^{-\frac{\alpha}{1+\alpha}} +\Omega_{ob}\biggr]
\end{multline}
\begin{figure}[ht]
\begin{center}
  \includegraphics[width=1.5\textwidth,natwidth=610,natheight=175]{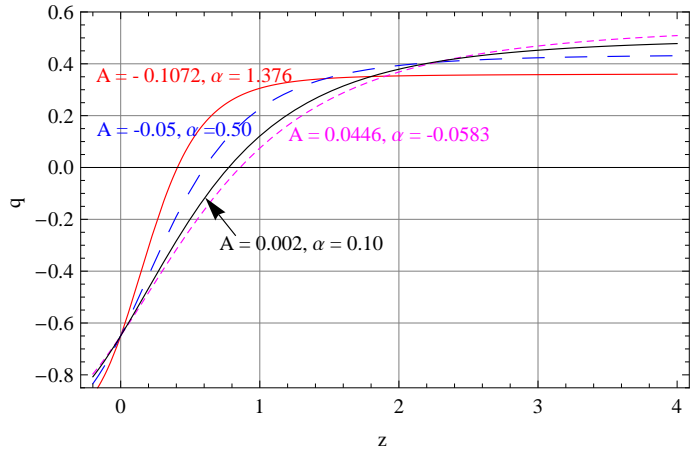}
  \caption{
  \small\emph{The variation of $q$ vs $z$  is shown in this
  figure.  Taking $\omega_{ode} = -1.05$,
   $\Omega_{ode} = 0.734$, $\Omega_{odm} = 0.222 $ and $\Omega_{ob}= 0.044$.  }\label{1}
    }
\end{center}
\end{figure}

In fig.  (9)  we plot the variation of the deceleration parameter
with redshift. It is found that the present deceleration parameter
at $z = 0$, \emph{i.e.}, $q_{0} = - 0.656$ which is independent of
the value of $A$ \& $\alpha$. This value is consistent with
observations ~\cite{mei} and follows from eq.  (27)  leading to
$(q)_{z=0} = \frac{1}{2}(1 + 3\omega_{ode}\Omega_{ode})$.
Deceleration \emph{flip} occurs at different values of $z$ (say
$z_{f}$) for different sets of $\alpha$ and $A$. For the values of
$A = - 0.1072 $ and $\alpha = 1.376$, \emph{flip} occurs at recent
past than that for other sets of the values of $A $ and $\alpha$,
in this case $z_{f} \approx 0.415$. It is also evident from fig.
(9) , $z_{f} \approx 0.60$ for $A = - 0.05$ \& $\alpha = 0.50$,
$z_{f} \approx 0.786$ for $A = 0.002$ \& $\alpha =0.100$ and
$z_{f} \approx 0.86$ for $A = 0.0446$ \& $\alpha = - 0.0583$. It
is further observed that for more negative values of $A$ flip
occurs at later time. However in all the cases an accelerating
universe emerges at low redshift value $(z < 0.8)$. We note that
at high redshift,  $q$ changes
 at slower rate with $z$ compared with high $z$($z > 2.0$). But
 it is evident at low value of $z$ \emph{i.e.}, in the recent past $q$ changes at
faster rate accommodating phantom like dynamics. It is also
observed in  fig. (9)  that as $A$ becomes more negative, the rate
of    decrease of $q$ is faster with $z$ accommodating an
accelerating universe.
 \vspace{0.5 cm}

\section{Observational Constraints on EoS Parameters}

Using eqs.  (23) to (25) we draw contours for the pairs of
parameters $(\alpha, A)$, $(\omega_{ode}, A)$ and $(\Omega_{odm},
A)$ in figs.    (1),  ( 2)  and (3)  respectively to study the
viability of the cosmological models with observational
predictions.

We note the following :

(i) For a given observational value of $\omega_{ode}$,
$\Omega_{odm}$ we draw ($\alpha$,$A$) contour in fig. (1). The
range of permitted values lies between $(-0.0903, 0.0446)$ for
$A$, $(- 0.0583, 1.012)$ for $\alpha$ at $68.3\%$ confidence
level. The range of permitted values lies between $(- 0.1074,
0.0984)$ for $A$, $(- 0.2241, 1.374)$ for $\alpha$ at $90\%$
confidence level. The range of permitted values lies between $(-
0.1166, 0.1417)$ for $A$, $(- 0.307, 1.585)$ for $\alpha$ at
$95.4\%$ confidence level.

(ii) For a given $\alpha$,  it is possible to draw contour between
$\omega_{ode}$ and $A$. Fig. (2)  is drawn with $\alpha = 0.05$,
$\Omega_{odm} = 0.22$. The range of permitted values of $A$ lies
between $(- 0.08172, 0.0068)$ and that for $\omega_{ode}$ between
$(- 0.9609, - 1.15)$ at $68.3 \%$ confidence level. The range of
permitted values of $A$ lies between $(- 0.0978, 0.0325)$ and that
for $\omega_{ode}$  between $(- 0.909, - 1.18)$ at $90 \%$
confidence level. The range of permitted values of $A$ lies
between $(- 0.1072, 0.0464)$ and that for $\omega_{ode}$  between
$(- 0.8771, - 1.194)$ at $95.4 \%$ confidence level.

(iii) For a given $\omega_{ode}$,  it is possible to draw contour
between $A$ and $\Omega_{odm}$.  In fig. (3) contours are   drawn
with $\alpha = 0.05$, $\omega_{ode} = -1.05$ using H-Z~\cite
{stern}, BAO~\cite {eisen}, CMB shift data ~\cite{jar}. We note
that the range of permitted values of $A$  lies between $( -
0.0583, - 0.0378)$ and that for $\Omega_{odm}$ lies between
$(0.1952, 0.2567)$ at $68.3 \%$ confidence level. We note that the
range of permitted values of $A$ between $(- 0.0616, - 0.0339)$
and that for $\Omega_{odm}$ lies between $(0.1846, 0.2699)$ at $90
\%$ confidence level. We note that the range of permitted values
of $A$ between $( - 0.0649, - 0.0319)$ and that for $\Omega_{odm}$
lies between $(0.1780, 0.2788)$ at $95.4 \%$ confidence level.

\section{Concluding Remarks}

    In this paper we have studied the dynamics of a flat FRW cosmology in
the framework of modified chaplygin gas and attempted  to obtain
the observational constraints of the  free parameters of the MCG
model employing  contour plot technique. We have also discussed,
in brief, the observational data, namely Stern data, measurement
of BAO peak parameters and CMB shift data to draw contours for the
permissible range of values of the pair of parameters ($\alpha ,
A$) and ($\omega_{ode} , A$) in figs. (1)  \& (2)
 to study the viability of cosmological
models \emph{vis-a-vis} observational results. We have taken some
of the permissible values of $A$ and $\alpha$ as $(-0.1072,10376);
(-0.05,0.50)$; $ (0.002,0.100)$ \& $(0.0446, -0.0583)$
respectively. However, from fig.  (3)  we determine  the
permissible range of $A$ which lies in the range ( $-0.0649$ to
$-0.0319$ ) at $95.4 \%$ confidence level. Thus the permitted
value of  $A$ for a viable cosmological model is very near to zero
from negative side. It may be of some interest to mention that MCG
model has an important application for the case, $A =
\frac{1}{3}$, pointing to a radiation dominated era. However our
analysis shows that the permissible range of values of $A$ do not
include $A = \frac{1}{3}$, which in our opinion, is quite feasible
because here we have focussed our attention to the late era only.
As pointed out in the introduction it may not
 be quite out of place to draw some correspondence to a fairly similar
 work  of Wang \emph{et al}~\cite{wang}  where they have studied the constraints on a
 decomposed GCG model described by dark matter interacting with
 inhomogeneous vacuum energy.  The first case ends with a very
 stringent constraint where $\alpha$ tends to vanishingly low
 value at 95\% confidence level. In the second case, however,
  the constraint is less restrictive such that $-0.15< \alpha < 0.25$.
  It may be interesting to point out that in our case with MCG model
  the range of $\alpha$ is  $- 0.1166 <  \alpha < 0.1417$ at 95.4\% confidence level.
   This weaker constraint matches favorably with the second case of their
    work. From the analysis presented here we summarize our
    findings as follows:

 (i). Taking the allowed  values of $A$  we studied the density dependence on $z$. As
expected we found from the fig. (4) that the density ( $\rho$)
decreases  with $z$. We have also calculated $\rho$ at $z = 0$ and
found  that
 $(1 - \Omega_{0b})$  is independent of $A$ and
$\alpha$  whereas  interestingly the variation $\frac{d\rho}{dz}$
is found to be a function of both $A$ and $\alpha$. We also note
that the universe is most dense for the set $(A = -0.1072, \alpha
= 1.376)$ \emph{i.e.} more negative value of $A$ leads to greater
energy density at early universe.
\vspace{0.5 cm}

(ii). The EoS parameter for $B = 0$ is $w(z) = A =
\frac{\omega_{0de}\Omega_{0de}}{1-\Omega_{0b}}$, which is negative
as $\omega_{0de}$ negative.  From fig. (5) , it is clear  that for
$A = 0.0446, \alpha = 0.0583$  a transition from ordinary matter
to dark energy dominated universe  is permitted as $w(z)$ transits
from positive to negative with a change in   $z$. For other sets
of values of $A$ and $\alpha$  here does not permit such
transition, since $w(z)$ always negative for any $z$.

From fig.  (5),  it is also evident that $w(z)$ is mostly flat at
high $z$ values ($z>2$) and very steep for low $z$ region. We note
 that at $z = 0$, $w(z)$ attains $-0.8$, which is  independent of
$A$ and $\alpha$, close to observational value.
 \vspace{0.5 cm}

(iii).  We have studied, in some detail the stability of our model
in section 3. The \emph{effective} speed of sound, $v_{s}^{2} <0$
at finite $z$ obtained here is similar to that obtained by Mei
~\cite{mei}
 where  $v_{s}^{2} <0$ at $z\rightarrow\infty$ .  This is  possibly
   due to the fact that Mei \emph{apriori}
chose a relation between $A$ and $\alpha$ ($A =
\frac{\alpha}{1+\alpha}$) unlike us. We further note from eq. (21)
that when, $A \geq 0$ the system is always stable but for $A < 0$
it depends on the relative magnitudes of $A$, $\omega (z)$ and
$\alpha$. Moreover it is further argued in the literature that
advent of $v_{s}^{2} <0$ signals the stage where structure
formation is likely to start.

\vspace{0.5 cm}

(iv).  It is evident from fig.  (7)  that normalised Hubble
parameter changes more sharply with $z$ for the set (A= -0.1072,
$\alpha$ =1.376) compared to other sets, where $A$ is less
negative in line with the fact that more negative $A$ values imply
a more accelerating model.
 \vspace{0.5 cm}

 (v). From fig. (8)  it is evident that for the set
($A = -0.1072,\alpha = 1.376$) density parameters change with $z$
at very fast rate than the other sets of the values of $A$ and
$\alpha$ considered here, \emph{i.e.}, density parameters change
faster for more negative values of $A$ admitting accelerating
universe. The gradual increasing density of dark energy is the
reason of the expansion of the universe which transit from
decelerating to accelerating phase.
\vspace{0.5 cm}

(vi). From fig. (9), it is observed that a  \emph{flip} in sign of
$q$ occurs at different values of $z_{f}$ for different sets of
$A$ and $\alpha$. It is also  observed that for more negative
values of $A$  \emph{flip} occurs at later time and the rate of
decrease is faster with $q$. However in all the cases \emph{flip}
occurs at low redshift value ($z < 0.8$) pointing to the fact that
accelerating phase is a late phenomenon in accordance with
observational predictions.
\vspace{0.5 cm}

Finally we note that in our model with more negative values of $A$
one obtains  \emph{(a)}  a  universe which is more dense
\emph{(b)} the density parameter changes faster with $z$
\emph{(c)} \emph{flip} occurs at a later time. Again from  fig.
(3), it is observed that the value of $A$ is negative and nearly
equal to zero which in agreement with the work of Fabris \emph{et
al}~\cite{fab3}. We have not considered  those values of $A$
because they do not change  the nature of evolution abruptly for
different parameters.

\vspace{0.5 cm}

To end a final remark may be in order. While the original
Chaplygin gas (CG) and its different variants are invoked to
explain the late time acceleration of the universe, recently CG
has found an important application \emph{vis a vis} the early
inflation ~\cite{campo}. In a recent communication Campo analysed
issues related to early inflation taking the GCG as source and has
attempted to constrain the value of the constant parameter
$\alpha$ with the help of recently released Planck data~\cite{par}
and in their study, for best fit the value of $\alpha $ comes out
to be, $\alpha =0.2578$. While unlike Campo here we are dealing
with MCG and moreover the formalism adopted by Campo's work is
quite different from what we have done here a cursory look at
fig.1, shows that, $\alpha = 0.25$ for $A=0$, the type of CG taken
by Campo. But we argue that this type of correspondence is
coincidental and need not be taken too far because while Campo
dealt with early era we have here discussed late era instead. The
analysis carried out by Campo may be extended in our model which
will be discussed elsewhere.  \vspace{0.5 cm}

\textbf{Acknowledgments} \vspace{0.5 cm}

SC acknowledges the award of a MRP from UGC, New Delhi  as also a
Twas Associateship award, Trieste. DP and SC acknowledge the local
hospitality of ITP, Beijing where a part of the work is done. B.
C. Paul also acknowledges the award of a MRP from UGC, New Delhi.
We also appreciate the comments and suggestions from the anonymous
referee, which has led to a significant improvement over the
previous version.
 \vspace{0.1 cm}

 \vspace{0.2 cm}


\begin{thebibliography}{35}
\bibitem{riess} Supernova Search Team collaboration, A.G. Riess \emph{et al}.,
Observational evidence from supernovae for an accelerating
universe and a cosmological constant, \emph{Astron. J.}
\textbf{116} (1998) 1009 [astro-ph/9805201] [SPIRES].

\bibitem{wmap} WMAP collaboration, D.N. Spergel \emph{et al}.,
First year Wilkinson Microwave
Anisotropy Probe (WMAP) observations: determination of
cosmological parameters, \emph{Astrophys. J. Suppl.} \textbf{148}
(2003) 175 [astro-ph/0302209] [SPIRES].

\bibitem{saif}
T. Padmanabhan, Dark energy: the cosmological challenge of the
millennium, \emph{Curr. Sci.} \textbf{88} (2005) 1057
[astro-ph/0411044] [SPIRES];  D. Panigrahi and S. Chatterjee,
\emph{JCAP} \textbf{10} (2011) 002; D. Panigrahi and S.
Chatterjee, \emph{Int. J. Mod. Phys.}\textbf{D21} (2012)12500791.

\bibitem{wen} S. Weinberg, \emph{Rev. Mod. Phys.} \textbf{61},(1989) 1.

\bibitem{quint} T. Chiba, N. Sugiyama and T. Nakamura, \emph{Mon. Not. Roy.
Astron. Soc.} \textbf{289}(1997) L5 [astro-ph/9704199]; R.
Caldwell, R. Dave and P. J. Steinhardt, \emph{Phys. Rev. Lett.}
\textbf{80} (1998) 1582 [astro-ph/9708069].

\bibitem{phan} R. R. Caldwell, \emph{Phys. Lett.}\textbf{B545}(2002)23
[astro-ph/9908168].

\bibitem{hol} M. Li, \emph{Phys. Lett.} \textbf{B603}(2004) 1 [hep-th/0403127].

\bibitem{string} L. McAllister and E. Silverstein, \emph{Gen. Rel. Grav.}
 \textbf{40} (2008) 565  [arXiv:0710.2951
 [hep-th]]; J. Polchinski, hep-th/0603249.

\bibitem{quantum} E. Elizalde, J. E. Lidsey, S. Nojiri and S. D. Odintsov,
\emph{Phys. Lett.} \textbf{B 574} (2003) 1
[hep-th/0307177].

\bibitem{star} A. Shafieloo, V. Sahni and A.A. Starobinsky, Is cosmic
acceleration slowing down?,
\emph{Phys. Rev. D} \textbf{80} (2009) 101301 [arXiv:0903.5141]
[SPIRES]; M. I. Wanas,  Dark energy: Is it of torsion origin?, in
Proc. First \emph{MEARIM}, eds. A. A. Hady and M. I. Wanas(2009),
arXiv: 1006.0476v1[gr-qc].

\bibitem{krasin} A. Krasinski, C. Hellaby, M.-N. C�el�erier
 and K. Bolejko, \emph{Gen. Rel. Grav.} \textbf{42}
 (2010) 2453 [arXiv:0903.4070] [SPIRES]; S. Chatterjee, \emph{JCAP}
  \textbf{03} (2011) 014 [arXiv:1012.1706]
 [SPIRES].

\bibitem{add}  K. Bamba, S. Capozziello, S. Nojiri and S. D.
Odintsov, \emph{Astrophys. Space Sci.} \textbf{342}(2012)155
[arXiv:1205.3421 [gr-qc]]; M. Li, X.D. Li, S. Wang and Y. Wang,
arXiv:1209.0922 [astro-ph.CO]; J. Yoo and Y. Watanabe, \emph{Int.
J. Mod. Phys.} \textbf{D21}(2012) 1230002,  arXiv:1212.4726
[astrph-ph.CO]; D. Panigrahi and S. Chatterjee, \emph{Gen.
Relativity and Grav.}\textbf{40} (2008)833.

\bibitem{lamda} E.J. Copeland, M. Sami and S. Tsujikawa, \emph{Int. J. Mod. Phys.} \textbf{D15}(2006)
1753 [hep-th/0603057]; D. Panigrahi and S. Chatterjee, \emph{Int.
J. Mod. Phys.} \textbf{A21}(2006) 6491.

\bibitem{sc}Isha Pahwa, Debajyoti Choudhury and
T.R.Seshadri, [arXiv:1104.1925v2[gr-qc]]; D. Panigrahi and S.
Chatterjee, \emph{Grav. and Cosm.} \textbf{17} (2011) 18; S.
Chatterjee and D.Panigrahi - DSU \textbf{335} (AIP) (2009) 1115,
gr-qc/0906.3847.

\bibitem{cg} A. Y. Kamenshchik, U. Moschella and V. Pasquir, \emph{Phys. Lett. } \textbf{B 511} (2001)  265; V. Gorini,
 A. Y. Kamenshchik, U. Moschella and V. Pasquir, \emph{Phys.
Rev. D} \textbf{69}(2004) 123512,[ gr-qc/0403062].



\bibitem{sand} H. Sandvik, M. Tegmark, M. Zaldarriaga, and I. Waga, \emph{Phys.
Rev. D} \textbf{69}(2004) 123524; V. Sahni, \emph{Lect. Notes
Phys.} \textbf{653}(2004) 141.


\bibitem{zhang} X. Zhang,  F. Q. Wu, and J. Zhang,  \emph{JCAP}\textbf{01}(2006)03.


\bibitem{ben} H. B. Benaoum, 2002, hep-th/ 0205140; U. Debnath, A. Banerjee
and S. Chakraborty,  \emph{Class. Quant. Grav.}\textbf{21}(2004)
5609.

\bibitem{fabris} J. C. Fabris, S. V. B. Goncalves and P. E.
de Souza, (2002) astro-ph/ 0207430; R. Jr. Colistete, J. C.
Fabris, S. V. B. Goncalves and P. E. de Souza,  \emph{Int. J. Mod.
Phys.} \textbf{D13}(2003) 669;  P. Thakur, S. Ghose and B. C.
Paul, \emph{Mon. Not. Roy. Astron. Soc.} \textbf{397} (2009) 1935.

\bibitem{deb} A. Dev, J. S. Alcaniz and D. Jain, \emph{Phys. Rev.}
 \textbf{D67}( 2003 ) 023515;
P. T. Silva and O. Bertolami,  \emph{Astrophys. J.}
\textbf{599}(2003) 829; A. Dev, D. Jain and J. S. Alcaniz,
\emph{Astron. Astrophys} \textbf{417}(2004) 847, astro-ph/
0311056.

\bibitem{alcaniz} J. S. Alcaniz, D. Jain and A. Dev, \emph{Phys.
Rev.} \textbf{D76}(2003)043514.

\bibitem{bento} M. C. Bento, O. Bertolami and A. A. Sen,  \emph{Phys.
Rev.}\textbf{D67} (2003) 063003.

\bibitem{cunha} J. V. Cunha, J. A. S. Lima and J. S. Alcaniz,
\emph{Phy. Rev.} \textbf{D69}(2004) 083501 ; astro-ph/0306319.

\bibitem{sahni} V. Sahni, T. D. Saini and A. A. Starobinsky and U.
Alam,  \emph{JETP Lett.} \textbf{77}(2003) 201; astro-ph /0210476.
\bibitem{lu} Jianbo Lu, Lixin Xu, Jiechao Li, Baorong Chang,
Yuanxing Gui and Hongya Liu, ar-Xiv: 1004.3364v1.
\bibitem{steinhardt} Paul J. Steinhardt,\emph{The Quintessential
Universe}: 20th.Texas Symposium(2001)(ed. J. C. Wheeler and H.
Marcel)\emph{American Institute of Physics}.


\bibitem{wang} Y. Wang, D. Wands, L. Xu, J De-Santiago and A. Hojjati,
 \emph{Phy. Rev.} \textbf{D87}(2013) 083503; ar-Xiv:
1301.5315 [astro-ph.CO].

\bibitem{fab3} J. C. Fabris, H. E. S. Velten, C. Ogouyandjou and
J. Tossa, \emph{Phys. Lett.} \textbf{B694} (2011) 289; ar-Xiv:
1007.10119(v1)[astro-ph.CO].

\bibitem{mei} Xue- Mei- Deng, \emph{Braz. J. Phys.}\textbf{41}(2011)
333.

\bibitem{fab} J. C. Fabris and J. Martin, \emph{Phys. Rev.} \textbf{D55}(1997)
5205.

\bibitem{stern} D Stern \emph{et al}, \emph{JCAP}\textbf{1002}(2010)008.

\bibitem{eisen} D. J. Eisenstein \emph{et al}, \emph{Astrophys. J.},
\textbf{633}(2005) 560.

\bibitem{jar}  Jarosik\emph{ et al}, \emph{ Astrophys. J. Suppl.}\textbf{192}(2011)
14.

\bibitem{campo}  S. del Campo, \emph{JCAP}\textbf{11}(2013)004 (and references therein).
14.

\bibitem{par}  Planck Collaboration, P.A.R. A de \emph{ et al}, Planck 2013 results XXII.
Constrained on inflation; ar-Xiv: 1303.5082.

\end{thebibliography}
\end{document}